\documentclass[12pt,preprint]{emulateapj}

\def\kms{$\rm km\, s^{-1}$}
\def\cm3{$\rm cm^{-3}$}

\def\n0{$\rm n_{0}$}
\def\B0{$\rm B_{0}$}

\def\mc{$\mu$m}

\def\L12{L$_{12\mu m}$~}
\def\F12{F$_{12\mu m}$~}

\def\fe2{[Fe\,{\sc ii}]}
\def\h2{H$_{2}$}

\shorttitle{CN molecular bands in AGNs}
\shortauthors{Riffel et al.}

\begin{document}


\title{The first detection of near-infrared CN bands in active galactic nuclei: 
signature of star formation}

\author{R. Riffel and M. G. Pastoriza}
\affil{Departamento de Astronomia, Universidade Federal do Rio Grande do Sul. 
	       Av. Bento Gon\c calves 9500, Caixa Postal 15051, CEP 91501-970,  Porto Alegre, RS, Brazil.}
\email{riffel@ufrgs.br}
\author{A. Rodr\'{\i}guez-Ardila\altaffilmark{1}}
\affil{Laborat\'{o}rio Nacional de Astrof\'{i}sica - Rua dos Estados Unidos 154,
Bairro das Na\c{c}\~{o}es.
CEP 37504-364, Itajub\'{a}, MG, Brazil.}

\altaffiltext{1}{ Visiting Astronomer at the Infrared   Telescope Facility, 
which is operated by the University of Hawaii under Cooperative Agreement no. 
NCC 5-538 with the National Aeronautics and Space Administration, Science 
Mission Directorate, Planetary Astronomy Program. }

\and 
\author{C. Maraston}
\affil{University of Oxford, Denys Wilkinson Building, Keble Road, Oxford, OX1 3RH, UK.}

\begin{abstract}

We present the first detection of the near-infrared CN absorption band in the nuclear 
spectra of active galactic nuclei (AGN). This feature is a  recent star formation tracer, 
being particularly strong in carbon stars. The equivalent width of the 
CN line correlates with that of the CO at 2.3 \mc\, as expected in stellar populations (SP)
with ages between $\sim$0.2 and $\sim$2\,Gyr. The presence of the 1.1\mc\ CN band 
in the spectra of the sources is taken as an unambiguous evidence of the presence 
of young/intermediate SP close to the central source of the AGN. 
Near-infrared bands can be powerful age indicators for star formation
connected to AGN, the understanding of which is crucial in the context
of galaxy formation and AGN feedback.

\end{abstract}


\section{Introduction}

It is widely known that circumnuclear star formation is commonly detected in active galactic nuclei
(AGN) \citep[e.g.][]{stu99,con02,thaisa05,sgp06}. In fact, in the last years, increasing 
observational evidence has confirmed that nuclear/circumnuclear  starbursts (SBs) coexists in objects harboring an AGN 
\citep[e.g.][]{msm94,id00,iman02,ara03,iw04}. For example, \citet{thaisa00} 
in a study of a sample of 20 Seyfert~2 (Sy~2) galaxies find out that about 50\% of the sources have 
young to intermediate-aged ($\sim$ 1 Gyr) nuclear starbursts and that 30\% of the galaxies
display a recent burst of star formation \citep[$t \rm < 500\, Myr$, results confirmed by][]{dhl01}. Currently, it is
thought that both the active nucleus and the starburst might be related to gas inflow, probably 
triggered by an axis-asymmetry perturbation like bars, mergers or tidal interactions 
\citep{sbf89,sbf90,maiolino97,ksp00,fathi06}. This gives support to the so-called AGN-starburst (AGN-SB)
connection \citep{ns88,terlevich90,heckman97,gd98,veilleux00,fmr00,hec04}. This connection, 
however, could be incidental, as many Seyferts  do not show any evidence of starburst activity 
\citep[e.g.][]{fil93} and optical spectroscopic studies of large samples do not indicate that 
starburst are more common in Seyferts than in normal galaxies \citep{pogge89}.  One difficulty in 
establishing the AGN-SB connection further is that tracing starbursts
reliably is difficult \citep{oliva95}. In the NIR region, except for a few indicators such as the 
methods based on the CO(2-0) first overtone or the Br$\gamma$ emission \citep[e.g.][]{orig93,oliva95},
the detection of spectral features allowing the identification and dating of young 
stellar populations (SP) in the inner few tens of parsecs of AGN remains difficult. 
To understand the physics of the above connection is particularly important for galaxy formation 
theories, as AGN activity is advocated as regulator of star formation in massive 
protogalaxies, the so-called AGN feedback phenomenon \citep{silk98,binney04,croton06,Bower06}.

In this line of thought, the evolutionary population synthesis
calculations presented by \citet[][hereafter M05]{maraston05}, by including
empirical spectra of carbon and oxygen rich stars \citep{lw00}, are 
able to foresee the presence of molecular features like CH, CN and C$\rm _2$. They 
arise in young/intermediate SP and their spectral signatures
are particularly enhanced in the NIR. Of particular importance
are the CN bands, which arise according to the models, from stars with ages in the range 
0.3$\leq t \leq$2 Gyr and are attributed to stars in the thermally 
pulsing asymptotic giant branch (TP-AGB) phase. It means that their 
detection can be taken as an unambiguous signature of the presence of young/intermediate 
SP in a well-constrained age, signaling the occurrence of SB activity. 
In the optical, CN bands are common in the spectra of globular clusters 
\citep[e.g.][]{smi89,coh99a,coh99b,smith06} but have been rarely reported in AGN. 
To our knowledge, the object showing the presence
of this feature is the Sy~2 galaxy NGC~7679, where \citet{gu01}
report extremely weak optical CN$\lambda$4200~\AA. However contrary to the NIR, optical CN reveals 
old SP rather than young/intermediate-age ones \citep[e.g.][]{thaisa00}. Here, we report the
first observation of the 1.1$\mu$m CN band predicted by the M05 models in 
a comprehensive sample of Seyfert galaxies, demonstrating the usefulness of the NIR spectral region 
as an important tool to investigate recent star formation episodes.   

This paper is structured as follows: in Section~\ref{dat} we describe the data. 
The results are presented in Section~\ref{results}. A discusion of the 
results is made in Section~\ref{disc}. Final remarks are given in Section~\ref{rem}.

\section{The data}\label{dat}

The galaxy spectra chosen for this work are the ones presented by 
\citet[][hereafter RRP06]{rrp06}.  The NIR spectra were obtained at the
NASA 3\,m Infrared Telescope Facility (IRTF). The SpeX spectrograph \citep{ray03}, was 
used in the short cross-dispersed mode. A 0.8''$\times$15'' slit was employed 
giving a spectral resolution of 360 \kms. For more details see RRP06.
A rapid inspection to the left panels 
of Figs. 1 to 8 and Fig. 12 of RRP06 shows that a significant 
fraction of objects display a deep prominent broad absorption feature starting at 
$\sim$1.1$\mu$m and extending up to 1.15$\mu$m in some sources (see, for
example, NGC\,1097, NGC\,1144 and NGC\,1614). In other objects such as
Mrk\,1066, NGC\,5953 and NGC\,7714, it is considerably narrower but still
deep and strong. Almost invariably, the galaxies displaying the 1.1$\mu$m
absorption also present strong CO 2.3$\mu$m bands and a {\it H-}continuum 
dominated by several absorption lines. The CaT at 0.8\,$\mu$m is also
prominent in these objects. Moreover, these sources were classified as 
Sy\,2/SB except Mrk\,334 and NGC\,1097, considered by RRP06 as Sy\,1. The invariably
association with CO bands and the CaT triplet allowed us to consider that 
the 1.1\,$\mu$m absorption could indeed be associated to stellar population 
although no previous observational report of this feature were found in the 
literature. 

Columns 2 and 3 of Table~\ref{data} list the names and AGN type, respectively, of the galaxies
where a clear evidence of the 1.1$\mu$m absorption was found. 

\begin{tiny}
\begin{table}
\renewcommand{\tabcolsep}{0.70mm}
\begin{center}
\caption{The sub-sample of RRP06 sample that shows the presence of the NIR CN molecular band. Some 
of the galactic properties and the equivalent widths of CN band and CO line are also presented.} \label{data}
\begin{tabular}{lccccccccc}
\tableline\tableline
ID  &  Galaxy       & Type   &      r	     & W$\rm _{CN}\,1.1\mu m$ & W$\rm _{CO}\,2.29\mu m$ \\
    &               &        &  (pc)$\rm ^a$ & (mag)		      & (mag) \\
(1) &  (2)          &  (3)   &   (4)	     &      (5) 	      & (6)  \\
\tableline
1   &	Mrk\,334     & Sy1   & 340   & 0.063$\pm$0.001  & 0.047$\pm$0.003 \\ 
2   &	NGC\,34      & SB/Sy2& 230   & 0.082$\pm$0.001  & 0.117$\pm$0.001  \\
3   &	NGC\,591     & Sy2   & 206   & 0.055$\pm$0.001  & 0.098$\pm$0.009 \\
4   &	Mrk\,573     & Sy2   & 267   & 0.046$\pm$0.002  & 0.057$\pm$0.005  \\
5   &   NGC\,1097    & Sy1   & 58    & 0.019$\pm$0.002  & 0.078$\pm$0.005  \\
6   &   NGC\,1144    & Sy2   & 447   & 0.062$\pm$0.002  & 0.077$\pm$0.009  \\
7   &	Mrk\,1066    & Sy2   & 186   & 0.045$\pm$0.001  & 0.097$\pm$0.007  \\
8   &	NGC\,1614    & SB    & 154   & 0.085$\pm$0.002  & 0.139$\pm$0.007  \\
9   &	NGC\,2110    & Sy2   & 121   & 0.066$\pm$0.001  & 0.033$\pm$0.001  \\
10  &	NGC\,3310    & SB    & 56    & 0.052$\pm$0.002  & 0.080$\pm$0.009 \\
11  &	NGC\,5929    & Sy2   & 193   & 0.050$\pm$0.002  & 0.082$\pm$0.009 \\
12  &	NGC\,5953    & Sy2   & 165   & 0.041$\pm$0.002  & 0.104$\pm$0.002  \\ 
13  &	NGC\,7682    & Sy2   & 179   & 0.036$\pm$0.001  & 0.079$\pm$0.009 \\
14  &   NGC\,7714    &H\,II  & 115   & 0.081$\pm$0.002  & 0.098$\pm$0.021  \\		    
\tableline
\tablenotetext{a}{Radius of the integrated region.}
\end{tabular}
\end{center}
\end{table}
\end{tiny}

\section{Results}\label{results}

The association of the 1.1\,$\mu$m absorption with the CN molecule was evident after
comparing our spectra with Fig. 15 of M05, where a grid of
synthetic spectra that include the effects of TP-AGB stars is shown. In these
models, the dominant feature in the interval 0.9$-$1.4\,$\mu$m is a strong absorption
centred at 1.1\,$\mu$m and attributed to CN. Models that do not include the 
TP-AGB population do not display the absorption. The M05 models also show that 
the strength of the CN absorption is a function of age and metallicity 
(Fig. 15 of M05). Other features present in the templates are also easily recognized in our data, including
the overall shape of the continuum emission and the CaT and CO bands, particularly
in the SB galaxies.

\begin{figure*}
\epsscale{0.80}
\plotone{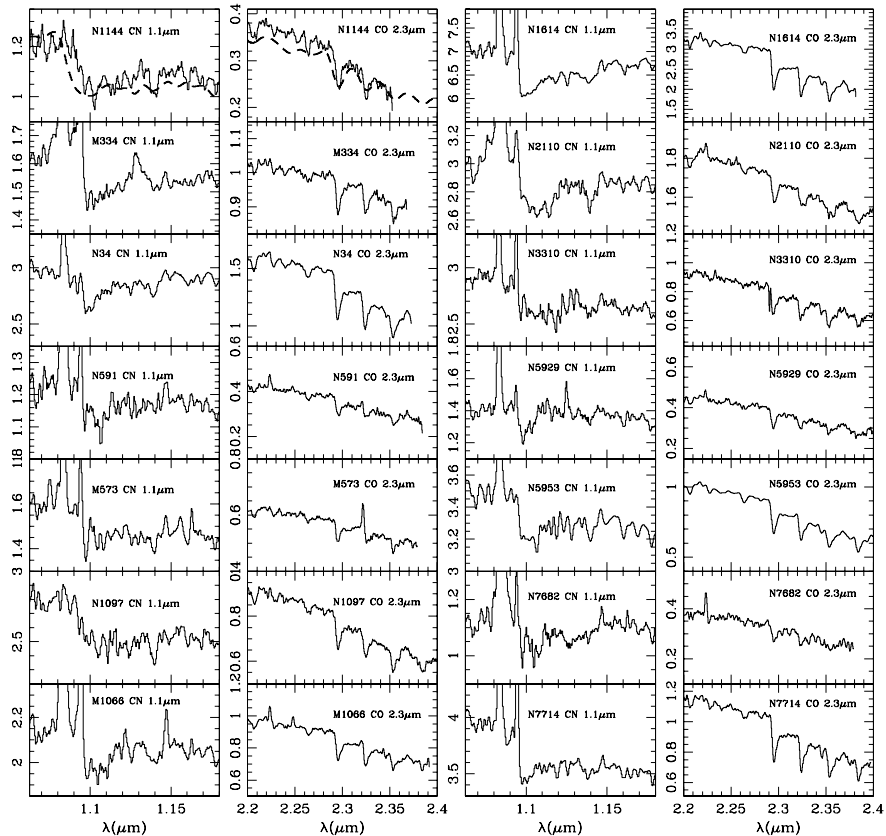}
\caption{Zoom around the 1.1\mc\ CN and 2.2\mc\ CO bands for the galaxies which clearly shows the 
presence of the 1.1\mc\ CN band. In the top left panel we show the predicted 1.1\mc\ CN band 
of a population with 1~Gyr and solar metallicity (M05) overploted at the NGC\,1144 
spectrum and the CN  molecular band lines observed in our sample of galaxies. 
The  ordinate is the flux in units of $\rm 10^{-15}\,erg\,cm^{-2}\,s^{-1}$. \label{cn}}
\end{figure*}

The upper left panel of Fig.~\ref{cn} shows a comparison of a 1~Gyr and 
solar metallicity template of  M05 with our
data, confirming that the 1.1\,$\mu$m absorption is due to CN, associated to
TP-AGB stars in the galaxies. No previous
report of the  CN1.1$\mu$m band in extragalactic objects was made before. Our result
points out to the presence of a significant young/intermediate stellar population, with ages between
0.3 and 2 Gyr in the objects listed in Tab.~1.

\begin{figure}
\epsscale{0.95}
\plotone{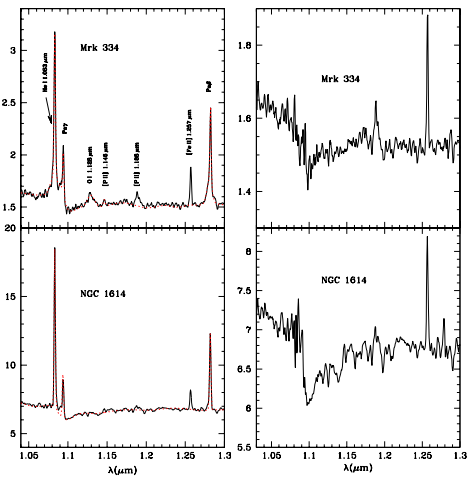}
\caption{Model for the emission lines around the CN absorption band. At the left side the full line is 
the spectrum of the galaxy and the modelled lines and continuum are represented by the dotted lines. At the right side
the CN band free from the emission spectra is presented.  The  ordinate is the flux in units 
of $\rm 10^{-15}\,erg\,cm^{-2}\,s^{-1}$.\label{model}}
\end{figure}

In order to properly quantify the strength of the CN 1.1\mc\ we
need to subtract the emission lines of He\,{\sc i} $\lambda$1.0830\mc\ 
and  Pa$\gamma$\,1.0938\,$\mu$m that affect the band. Therefore, we have modelled these emission 
lines constraining the Pa$\gamma$ width and flux to those of Pa$\beta$\,1.2820\,$\mu$m
using the calculations by \citet{hs87} for their intrinsic line ratio values.
For the He\,{\sc i}, we have adjusted a gaussian to the emission line profile. It is a reliable
approximation because the helium emission line is located near the blue end of the CN absorption
band. As for the He\,{\sc i}, in the spectrum of Mrk\,334 we have adjusted a gaussian to the O\,{\sc i} 
$\lambda$1.1287\mc\ emission line. The LINER 
software \citep{pow93} was used for this purpose. A sample of the results of this process 
is presented in Fig.~\ref{model}.

With a clear CN band, free of contamination of the emission lines, we proceed to measure
the equivalent width (W) of the CN band. The continuum was linearly adjusted using the points between 
1.0445\mc\ and 1.0580\mc\ as representative of the continuum at the left side of the CN band and for the right side
we defined the points between 1.2160\mc\ and 1.2385\mc.  The band widths were defined using the Maraston models,
as being between 1.0780\mc\ and 1.1120\mc.  We also include, for comparison, the W of the CO 2.2980 \mc\ line of 
the 2.3\mc\ CO band, measured from the galaxy spectra, taking as the line width the points 2.2860\mc\ and 2.3100\,\mc\ 
for the band width and a continuum defined as a spline using points free from emission/absorption lines in 
the interval between 2.2350\mc\ and 2.3690\mc. The W$\rm _{CN}$ and W$\rm _{CO}$ measured values are listed 
in column~7 and 8 of Tab.~\ref{data}. Here the LINER software was also used.

\section{Discussion }\label{disc}

The NIR CN absorption band is particularly enhanced in carbon
stars, where there is residual carbon that is not binded with oxygen
in CO molecules. As carbon stars are only produced after the third
convective dredge-up along the TP-AGB phase \citep[e.g.][]{ir83}, their
features are a clear indicator of SP rich in these
stars. This constrains the age of such SP to be
within the narrow range 0.3 to 2\,Gyr (M05).

\begin{figure}
\epsscale{0.90}
\plotone{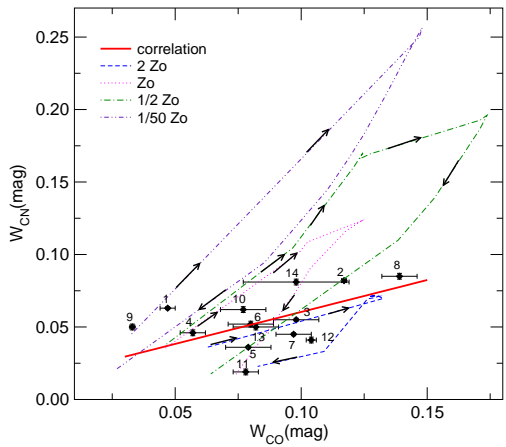}
\caption{CN equivalent width {\it versus} CO equivalent width. The observed 
values are represented by full circles. Numbers identify the sources
according to column~1 of Tab~\ref{data}. Models are indicated on the labels. The arrows 
indicate the age sequence of the models ranging from 0.2 Gyr to 3 Gyr. \label{cnco}}
\end{figure}

In order to check if our observations are actually consistent with
this claim, we compare the observed strength of CN and CO with the Maraston's
model predictions.  Note that the CO alone would not be a useful age indicator because
this band is not only strong in TP-AGB stars or red-supergiants, but
also in old RGB stars \citep[e.g.][]{ivanov00,iah04}. Fig.~\ref{cnco} plot the values of simple stellar
populations (SSPs) with ages between 0.2 and 3 Gyr and metallicities
from 1/50 to twice solar.

Fig.~\ref{cnco} confirms that the observed values are indeed consistent with
the model predictions in this age range, and that the strengths of the
two bands are correlated  (R$^2$= 0.66).  
The CO band is also present in TP-AGB spectra (cfr Fig.~16 of M05), which explains
the correlation between the W of CN and CO.
Interestingly, the data can be
separated into two main groups according to the chemical
composition. One group (NGC\,34, Mrk\,573,  NGC\,1144, 
NGC\,3310, NGC\,7682 and NGC 7714) has the light dominated by stars
with solar/half solar metallicity. A second group (NGC\,591, NGC\,1097, 
Mrk\,1066, NGC\,5929 and NGC 5953) harbors SP with
high metallicity, and just two
objects are consistent with low metallicity. A detailed analysis of
the ages and metallicities of our sample is postponed to another paper
(Riffel et al., {\it in preparation}).

Optical stellar population synthesis carried out by \citet{raimann03} in a sample of Seyfert galaxies,
four of them common to our sample (Mrk\,573, Mrk\,1066, NGC\,2110 and NGC\,5929) reveals that old 
population (10 Gyr) dominates the inner few hundred parsecs. The integrated 
regions of our NIR spectra are similar to that of \citet{raimann03}. As shown by our results, 
for these same objects the NIR spectra display unambiguous signature of young/intermediate SP 
with ages between 0.3 and 2 Gyr. It means that the J-band provides complementary information not 
detected in the optical. An additional advantage of the CN is that 
it is located near the center of the $J$-band allowing its observation in a wide range of 
redshifts ($z \lesssim\,0.18$). The CO band instead is located near the red end of
the K-band thereby allowing only observations of sources with
redshifts smaller than 0.03.

\section{Final Remarks}  \label{rem}

We present the first detection of the $\rm 1.1 \mu m$ CN band in the inner 300 pc of an AGN. The  
presence of the CN at 1.1\mc\ in the spectrum can be taken as an unambiguous 
evidence of recent star formation, in particular it is associated with the presence
of bright carbon stars. We have compared the observed values with predictions of SP including the contribution of carbon stars. We find a nice
consistency of observed and predicted values for ages around 1
Gyr. This supports the claim that the NIR CN  band is suggestive of
young/intermediate age SP. Its use as age indicator
is particularly adapt to AGN-hosts, where dust reddening can
complicate seriously the use of optical bands. Moreover, since these
molecules are enhanced in SP spanning a narrow age
range, their detection is relatively robust against the age/metallicity
degeneracy.

\acknowledgments

This paper was partially supported by the Brazilian funding agency
CNPq(304077/77-1) to ARA. CM is a Marie Curie Excellence Team Leader and holds grant
MEXT-CT-2006-l 042754 of the Training and Mobility of Researchers
programme financed by the European Community.


\begin{thebibliography}{}
\bibitem[Binney(2004)]{binney04} Binney, J. 2004, MNRAS, 347, 1093.
\bibitem[Bower et al.(2006)]{Bower06} Bower, R. G.; Benson, A. J.; Malbon, R.; Helly, J. C.; Frenk, C. S.; Baugh, C. M.; Cole, S.; Lacey, C. G., 2006, MNRAS, 370, 645.
\bibitem[Cohen(1999a)]{coh99a} Cohen, J. G., 1999, AJ, 117, 2434.
\bibitem[Cohen(1999b)]{coh99b} Cohen, J. G., 1999, AJ, 117, 2428.
\bibitem[Contini et al.(2002)]{con02} Contini, M., Radovich, M., Rafanelli, P., Richter, G. M., 2002, ApJ, 572, 124.
\bibitem[Croton et al.(2006)]{croton06} Croton, Darren J.; Springel, Volker; White, Simon D. M.; De Lucia, G.; Frenk, C. S.; Gao, L.; Jenkins, A.; Kauffmann, G.; Navarro, J. F.; Yoshida, N., 2006, MNRAS, 365, 11.
\bibitem[Fathi et al.(2006)]{fathi06} Fathi, K., Storchi-Bergmann, T., Riffel, R. A., Winge, C., Axon, D. J., Robinson, A., Capetti, A., Marconi, A., 2006, ApJ, 641, 25.
\bibitem[Filippenko et al.(1993)]{fil93} Filippenko, A. V., Ho, L. C., Sargent, W. L. W., 1993, ApJ, 410, 75.
\bibitem[Ferrarese \& Merritt(2000)]{fmr00} Ferrarese, L., \& Merritt, D., 2000, ApJ, 539, L9	
\bibitem[Gonzalez-Delgado et al.(1998)]{gd98}Gonzalez Delgado R.M., Heckman T., Leitherer C., Meurer G., Krolik J., Wilson A.S., Kinney A., Koratkar A., 1998, ApJ, 505, 174.
\bibitem[González-Delgado et al.(2001)]{dhl01} González-Delgado, R. M., Heckman, T. \& Leitherer, C., 2001, ApJ, 546, 845
\bibitem[Gu et al.(2001)]{gu01} Gu, Q. S.; Huang, J. H.; de Diego, J. A.; Dultzin-Hacyan, D.; Lei, S. J.; Benitez, E., 2001, A\&A, 374, 932.
\bibitem[Heckman et al. (1997)]{heckman97} Heckman T.M., Gonzalez-Delgado R.M., Leitherer C., Meurer G.R., Krolik J., Wilson A.S., Koratkar A., Kinney A., 1997, ApJ, 482, 114.
\bibitem[Heckman(2004)]{hec04}Heckman, T. M., 2004, cbhg.symp, 358.
\bibitem[Hummer \& Storey(1987)]{hs87} Hummer, D. G., \& Storey, P. J. 1987, MNRAS, 224, 801
\bibitem[Iben \& Renzini(1983)]{ir83} Iben \& Renzini 1983, ARA\&A, 21, 271.
\bibitem[Imanishi \& Dudley(2000)]{id00} Imanishi, M., Dudley, C. C., 2000, ApJ, 545, 701.
\bibitem[Imanishi(2002)]{iman02} Imanishi, M., 2002, ApJ, 569, 44.
\bibitem[Imanishi \& Alonso-Herrero(2004)]{iah04} Imanishi, M.; Alonso-Herrero, A., 2004, ApJ, 614, 122.
\bibitem[Imanishi \& Wada(2004)]{iw04} Imanishi, M. \& Wada, K., 2004, ApJ, 617, 214.
\bibitem[Ivanov et al.(2000)]{ivanov00} Ivanov, V. D.; Rieke, G. H.; Groppi, C. E.; Alonso-Herrero, A.; Rieke, M. J.; Engelbracht, C. W., 2000, ApJ, 545, 190.
\bibitem[Knapen et al.(2000)]{ksp00} Knapen, J. H.; Shlosman, I. \& Peletier, R. F., 2000, ApJ, 529, 93.
\bibitem[Lancon \& Wood(2000)]{lw00} Lancon \& Wood 2000, A\&AS, 146, 217.
\bibitem[Maiolino et al.(1997)]{maiolino97} Maiolino, R., Ruiz, M., Rieke, G. H., Papadopoulos, P., 1997, ApJ, 485, 552.
\bibitem[Maraston(2005)]{maraston05} Maraston, C., 2005, MNRAS, 362, 799. M05
\bibitem[Mizutani et al.(1994)]{msm94} Mizutani, K., Suto, H., Maihara, T., 1994, ApJ, 421, 475
\bibitem[Norman \& Scoville(1988)]{ns88} Norman C., Scoville N., 1988, ApJ, 332, 124.
\bibitem[Oliva et al.(1995)]{oliva95} Oliva, E.; Origlia, L.; Kotilainen, J. K.; Moorwood, A. F. M.,  1995, A\&A, 301, 55.
\bibitem[Origlia et al.(1993)]{orig93}  Origlia, L.; Moorwood, A. F. M.; Oliva, E., 1993,A\&A, 280, 536.
\bibitem[Pogge(1989)]{pogge89} Pogge, R. W., 1989, ApJS, 71, 433.
\bibitem[Pogge \& Owen(1993)]{pow93} Pogge, R. W., \& Owen, J. M. 1993, OSU Internal Report 93-01
\bibitem[Raimann et al.(2003)]{raimann03} Raimann, D., Storchi-Bergmann, T., González Delgado, R. M., Cid Fernandes, R., Heckman, T., Leitherer, C., Schmitt, H., 2003, MNRAS, 339, 772.
\bibitem[Rayner et al.(2003)]{ray03} Rayner, J. T., Toomey, D. W., Onaka, P. M., Denault, A. J., Stahlberger, W. E., Vacca, W. D., Cushing, M. C., \& Wang, S. 2003, PASP, 155, 362
\bibitem[Riffel et al.(2006)]{rrp06} Riffel, R.; Rodr\'{\i}guez-Ardila, A.; Pastoriza, M. G., 2006, A\&A 457, 61. RRP06
\bibitem[Rodr\'{\i}guez-Ardila \& Viegas(2003)]{ara03} Rodr\'{\i}guez-Ardila, A.; Viegas, S. M., 2003, MNRAS, 340, 33
\bibitem[Shi et al.(2006)]{sgp06} Shi, L.; Gu, Q. S. \& Peng, Z. X., 2006, A\&A, 450, 15. 
\bibitem[Shlosman et al.(1989)]{sbf89}Shlosman, I., Frank, J. \& Begelman, M. C., 1989, Natur, 338, 45.
\bibitem[Shlosman et al.(1990)]{sbf90} Shlosman, I., Begelman, M. C. \& Frank, J., 1990, Natur, 345, 679.
\bibitem[Silk \& Rees(1998)]{silk98}Silk, J., Rees, M. J. 1998, A\&A, 331,1
\bibitem[Smith \& Briley(2006)]{smith06}Smith, Graeme H.; Briley, Michael M., 2006, PASP, 118, 740.
\bibitem[Smith et al.(1989)]{smi89} Smith, G. H., Bell, R. A. \& Hesser, J. E., 1989, PASP, 101, 1083.
\bibitem[Sturm et al.(1999)]{stu99} Sturm, E., et al., 1999, ApJ, 512, 197.
\bibitem[Storchi-Bergmann et al.(2000)]{thaisa00} Storchi-Bergmann, T., Raimann, D., Bica, E. L. D., Fraquelli, H. A.,  2000, ApJ, 544, 747
\bibitem[Storchi-Bergmann et al.(2005)]{thaisa05} Storchi-Bergmann, T.; Nemmen, R. S.; Spinelli, P. F.; Eracleous, M.; Wilson, A. S.; Filippenko, A. V.; Livio, M., 2005, ApJ, 624L, 13.
\bibitem[Terlevich et al.(1990)]{terlevich90} Terlevich E., Diaz A.I.,  Terlevich R., 1990, MNRAS, 242, 271.
\bibitem[Veilleux(2000)]{veilleux00} Veilleux S., 2000, 2001, sgnf.conf, 88.
\end{thebibliography}
\end{document}